\documentclass[preprint,12pt]{aastex}

\shorttitle{Francis et al.}
\shortauthors{Elliptical Galaxies at z=2.38}

\begin{document}

\title{A Pair of Compact Red Galaxies at Redshift 2.38,
Immersed in a 100 kpc Scale Ly$\alpha$ Nebula\altaffilmark{1}}

\author{
Paul J. Francis\altaffilmark{2} \altaffilmark{3},
Gerard M. Williger\altaffilmark{4}, 
Nicholas R. Collins\altaffilmark{5}, 
Povilas Palunas\altaffilmark{6}, 
Eliot M. Malumuth\altaffilmark{5}, 
Bruce E. Woodgate\altaffilmark{7}, 
Harry I. Teplitz\altaffilmark{4}, 
Alain Smette\altaffilmark{4}, 
Ralph S. Sutherland\altaffilmark{2}
Anthony C. Danks\altaffilmark{8}, 
Robert S. Hill\altaffilmark{5}, 
Donald Lindler\altaffilmark{9}, 
Randy A. Kimble\altaffilmark{7}, 
Sara R. Heap\altaffilmark{7},
John B. Hutchings\altaffilmark{10}
}

\altaffiltext{1}{Based upon observations with the Anglo-Australian
Telescope, the Cerro Tololo Blanco
Telescope, observations taken with the ESO New Technology Telescope (NTT) at 
the La Silla
Observatory under program-ID No.63.0-0291(A), and observations taken with 
the NASA/ESA Hubble
Space Telescope, obtained at the Space Telescope Science Institute,
which is operated by the Association of Universities for Research in
Astronomy, Inc., under NASA contract No. NASS-26555}
\altaffiltext{2}{Research School of Astronomy and Astrophysics, The 
Australian National University, Canberra, ACT 0200, Australia}
\altaffiltext{3}{Joint appointment with the Department of Physics, Faculty
of Science.}
\altaffiltext{4}{NOAO, Code 681, NASA Goddard Space Flight Center, 
Greenbelt, MD 20771}
\altaffiltext{5}{Raytheon ITSS, Code 681, Goddard Space Flight Center, 
Greenbelt, MD 20771}
\altaffiltext{6}{Catholic University of America, Goddard Space Flight 
Center, Code 681, Greenbelt, MD 20771}
\altaffiltext{7}{NASA Goddard Space Flight Center, Code 681, Greenbelt, 
MD 20771}
\altaffiltext{8}{Raytheon RPSC, Code 681, Goddard Space Flight Center, 
Greenbelt, MD 20771}
\altaffiltext{9}{Advanced Computer Concepts, Inc., Goddard Space Flight 
Center, Code 681, Greenbelt, MD 20771}
\altaffiltext{10}{Dominion Astrophysical Observatory, Victoria, BC 
V8X 4M6, Canada}

\begin{abstract}

We present Hubble Space Telescope (HST) and ground-based observations
of a pair of galaxies at redshift 2.38, which are collectively known as
2142$-$4420 B1 \citep{F96}. The two galaxies are both luminous extremely red
objects (EROs), separated by 
0.8\arcsec . They are embedded within a 100 kpc scale diffuse Ly$\alpha$ 
nebula (or blob) of luminosity $\sim 10^{44}{\rm \ erg\ s}^{-1}$.

The radial profiles and colors of both red objects are most naturally
explained if they are 
young elliptical galaxies: the most distant yet found. It is not,
however, possible to rule out a model in which they are abnormally 
compact, extremely dusty starbursting disk galaxies. If they are elliptical
galaxies, their stellar 
populations have inferred masses of $ \sim 10^{11} M_{\sun}$ and ages 
of $\sim 7 \times 10^{8}$ years. Both galaxies have color gradients: their 
centers are significantly bluer than their outer regions. The surface
brightness of both galaxies is roughly an order of magnitude greater than
would be predicted by the Kormendy relation.
A chain of diffuse star formation extending 1\arcsec\ from the galaxies
may be evidence that they are interacting or merging. 

The Ly$\alpha$ nebula surrounding the galaxies shows apparent velocity
substructure of amplitude $\sim 700 {\rm \ km\ s}^{-1}$. We propose that
the Ly$\alpha$ emission from this nebula may be produced by fast shocks,
powered either by a galactic superwind or by the release of gravitational
potential energy. 

\end{abstract}


\section{Introduction\label{intro}}

Two of the most enigmatic types of high redshift galaxy are the
extremely red objects \citep[EROs, eg.][]{hu94,tho99}, and the
Ly$\alpha$ blobs \citep[eg.][]{lo91,F96,kee99,ste00,roc00,kob00}. 

EROs are
defined as having extremely red observed-frame optical/near-IR colors, and may
be dusty starburst galaxies or AGN \citep[eg.][]{sma99,hug98}, or  
spheroidal galaxies \citep[eg.][]{cim99,dun96,soi99,mor00}. 

The Ly$\alpha$
blobs are characterized by 100 kpc scale low surface brightness Ly$\alpha$ 
nebulae, with total Ly$\alpha$ luminosities $>10^{43}{\rm erg\ s}^{-1}$. All 
Ly$\alpha$ blobs identified to date appear to lie in proto-cluster 
environments. It
has been proposed that these blobs are a form of AGN, that they are
associated with cooling flows \citep{ste00,hai00} or that they are generated
by galaxy superwinds \citep{tan00}.

Can a galaxy be both an ERO and a Ly$\alpha$ blob?
Many high redshift radio galaxies contain ERO nuclei embedded within
dense gaseous regions, which are often associated with extended diffuse 
Ly$\alpha$ emission \citep[eg.][]{mcc93,pen97,ivi00,car97,bin00,bic00}.

In this paper we present a detailed study of a {\em radio quiet} Ly$\alpha$
blob that contains two compact EROs. \citet{F96} identified 
2142$-$4420 B1, which is a 
luminous Ly$\alpha$ and \ion{C}{4} emitting source at redshift $z=2.38$. 
The enormous Ly$\alpha$ luminosity of this source  
comes from an extended $\sim 100$ kpc low surface brightness nebula.
Ground-based near-IR observations showed an unresolved, luminous, extremely 
red source lying within this blob. The whole system 
appears to lie within a cluster of Lyman-limit QSO absorption-line
systems \citep{F00}. The optical properties of this source closely
resemble those of many high redshift radio galaxies, but \citet{F96}
obtained an upper limit on its radio flux of 0.27 mJy at 1344 and 2378
MHz.

We describe our observations in Section~\ref{obs},
and show the results in Section~\ref{results}. We then discuss the
nature of the ERO (Section~\ref{continuum}) and of the Ly$\alpha$
blob surrounding it (Section~\ref{lines}). In Section~\ref{conclusions} we 
conclude
that this object is probably a pair of young merging elliptical galaxies,
though a compact dusty starburst model is hard to completely exclude.
We further conclude that
the Ly$\alpha$ nebula surrounding them is either photoionized
by an AGN or shock excited by the energy of gravitational 
infall or of a superwind.

We assume a flat universe with $H_0 = 70 {\rm km\ s}^{-1}{\rm Mpc}^{-1}$, 
$\Omega_m = 0.3$ and $\Omega_{\Lambda} = 0.7$. 
At $z=2.38$, given this cosmology and redshift, 
one arcsecond corresponds to $8.0$ proper Kpc, and the 
luminosity distance is $18.8\ {\rm Gpc}$.

\section{Observations\label{obs}}

Our previous observations of this object can be found in \citet{FH93,F96,
F97} and \citet{F00}. The observation log is in Table~\ref{obstab}. HST 
imaging was used to determine the morphology and colors
of B1. Ground-based imaging measured its integrated colors and 
diffuse Ly$\alpha$ emission. Slitless spectroscopy and Fabry-Perot imaging
was used to constrain the velocity structure.

In addition to our own observations, we searched the ROSAT All Sky Survey
\citep{vog99} 
for data on this object. There is only a 308 sec exposure toward the field, 
giving a 90\% upper limit X-ray count rate of 0.0075 counts sec$^{-1}$.   
Assuming a Galactic column density of $2.65 \times 10^{20}$ cm$^{-2}$
along this line of sight
(which is the weighted mean of the four nearest 21 cm measurements) and a 
power law spectrum with an energy index of 1.0, we obtain a flux
$f_E < 2.6\times 10^{-5}$ E$^{-2}$ photons cm$^{-2}$ sec$^{-1}$ keV$^{-1}$
where E is the energy in keV.

\begin{scriptsize}
\begin{deluxetable}{lccclc}
\scriptsize
\tablewidth{0pt}
\tablecolumns{4}
\tablecaption{Observations \label{obstab}}
\tablehead{
\colhead{Instrument} &
    \colhead{Filter} &
     \colhead{Rest-frame $z=2.38$} &
            \colhead{Exposure} &
                                   \colhead{Date} &
                                                    \colhead{Notes} \\
\colhead{} &  \colhead{} & \colhead{wavelength (nm)}  
 & \colhead{(s)} & \colhead{ } & \colhead{ }
}
\startdata
HST/WFPC2   & F410M   & $121 \pm 2.1$ & 67200 & 1999 Aug 27---Sep 8  & \\
HST/WFPC2   & F450W   & $131 \pm 13$  & 19400 & 1999 Aug 31---Sep 2  & \\
HST/WFPC2   & F702W   & $207 \pm 21$  & 14400 & 1999 Aug 23          & \\
HST/NICMOS  & F110W   & $325 \pm 87$ & 5120 & 1998 Jun 27          & \\
HST/NICMOS  & F160W   & $471 \pm 54$ & 10112 & 1998 Oct 30 & \tablenotemark{a} \\
HST/NICMOS  & F164N   & $487 \pm 2.5$ & 10236 & 1997 Oct 30 & \tablenotemark{a} \\
HST/STIS    & Clear   &  &  3120 & 1998 Oct 25---27     & \tablenotemark{b} \\
HST/STIS    & G430L   &  & 17000 & 1998 Oct 25---27     & \tablenotemark{c} \\
NTT/SOFI    & $J$     & $369 \pm 43$ &  3540 & 1999 Aug 16          & \\
NTT/SOFI    & $H$     & $489 \pm 44$ &  4740 & 1999 Aug 16          & \\
NTT/SOFI    & $K_s$   & $640 \pm 40$ &  3480 & 1999 Aug 16          & \\
CTIO/Mosaic & 4107\AA & $121.5 \pm 0.8$ & 18000 & 1999 Aug 7           & \\
CTIO/Mosaic & $U$     & $108 \pm 10$ &  9000 & 1999 Aug 8           & \\
CTIO/Mosaic & $B$     & $130 \pm 14$ &  7200 & 1999 Aug 7---8       & \\
CTIO/Mosaic & $V$     & $163 \pm 13$ &  7800 & 1999 Aug 7---8       & \\
CTIO/Mosaic & $R$     & $207 \pm 32$ &  2400 & 1999 Aug 8           & \\
CTIO/Mosaic & $I$     & $260 \pm 35$ &  4800 & 1999 Aug 8           & \\
AAT/TTF     & 4110\AA & $121.6 \pm 0.6$ & 5400 & 1999 Sep 11 & \tablenotemark{d} \\
AAT/TTF     & 4114\AA ? & $121.7 \pm 0.6$ & 5400 & 1999 Sep 11 & \tablenotemark{d} \\
AAT/TTF     & 4118\AA & $4121.8 \pm 0.6$ & 9000 & 1999 Sep 12 & \tablenotemark{d} \\
\enddata
\tablenotetext{a}{Not taken during a NIC3 focus campaign. Image
quality degraded, and image partially vignetted}
\tablenotetext{b}{Taken to allow registration of the slitless spectroscopy}
\tablenotetext{c}{Slitless spectroscopy}
\tablenotetext{d}{Wavelength calibration uncertain}
\end{deluxetable}
\end{scriptsize}

\subsection{WFPC2 Imaging\label{wfpc}}

The cluster region was observed for 38 orbits with the Wide Field and
Planetary Camera II \citep[WFPC2, ][]{tra94} on the Hubble Space Telescope. 
Galaxy B1 was placed on the
WF3 chip. Exposures were dithered on a sub-pixel grid, to allow reconstruction
of a better sampled image.
Broad-band images were obtained through the F450W and F702W filters.
Relatively narrow-band images were taken through the F410M filter, which 
matches the wavelength of Ly$\alpha$ at the cluster wavelength. 
The F410M images suffer from a very low surface brightness, time-varying 
mottling. We hypothesise that this is caused by scattered Earth-light.

The images were pipeline processed, then cosmic rays were removed and
the frames co-added using the drizzle
algorithm \citep{fru98} to minimise undersampling. 

\subsection{STIS Slitless Spectroscopy\label{stis}}

The Space Telescope Imaging Spectrograph \citep[STIS, ][]{woo98} observations 
(pairs of 
slitless spectra and images) were taken at two different orientation
angles separated by 11.5 deg, in an attempt to separate slitless
spectra from objects aligned along the same row for a given
angle.  For both the direct clear-aperture and the dispersed images, the 
CCD was binned $2\times 2$, producing a
plate scale of $0.1$ arcsec/pixel, and spectral resolution
$\sim 11$ \AA\ for point sources.  There were 20 exposures each for the
direct and dispersed images, with dithering of $\sim 1.4$ arcsec between
each exposure in order to minimize the effects of cosmic rays and hot pixels.
The direct and dispersed images were combined using the STIS GTO
team software CALSTIS \citep{lin98,gar98}.  
The spectra were extracted using both the {\sc stis\_extract}
\citep{che99} and {\sc slwidget} \citep{lin00} software. Results from both 
were consistent.

\subsection{NICMOS Imaging\label{nicmos}}

The Near Infrared Camera and Multi Object Spectrometer
\citep[NICMOS, ][]{tho98} observations were made with Camera 3, through 
the F110W, F160W and F164N filters.  All
exposures were dithered for each filter over 4 (or 8 in the case of F160W)
positions spanning a $2\times 2$ arcsec box, to minimize the effects of bad
pixels. The F160W and F164N observations are affected by
vignetting from the field offset mirror, and were taken before the
focus became optimal. The NICMOS data were reduced using the same methods 
as for SOFI (Section~\ref{irim}).  The data were also reduced 
using NICRED \citep{mcl97}; there is little difference between final images 
reduced by the two methods. A comparison of broadband SOFI and NICMOS data 
show the photometry to be consistent between the two, with the exception of a
vignetted band along one end of the chip in the F160W and F164W images. 

\subsection{NTT Imaging\label{irim}}

Near-IR photometry was obtained with the SOFI camera on 
the ESO New Technology Telescope \citep[NTT, ][]{lid00}  . Conditions were 
photometric,
but seeing was $\sim$ 1.1\arcsec\ and the telescope was subject to 
considerable wind shake. The data were reduced using a modified
version of IRAF\footnote{IRAF is distributed by the 
National Optical Astronomy Observatories, which is operated by the
Association of Universities for Research in Astronomy, Inc. (AURA)
under cooperative agreement with the National Science Foundation.}
scripts written by Peter McGregor.  NICMOS standards \citep{per98} were used.

\subsection{Cerro Tololo 4m Imaging\label{ctio}}

We obtained broad- and narrow-band (Ly$\alpha$) imaging of this field with the
MOSAIC camera \citep{mul98}, at the prime focus of the Cerro Tololo
Inter-American Observatory (CTIO) 4m Blanco Telescope. The narrow-band 
images were taken through a
specially purchased filter, which working in the f/2.8 beam has
a bandpass of 54\AA , centered at 4107\AA . Conditions were mostly
photometric, with seeing $\sim 1.2$\arcsec.

\subsection{Anglo-Australian Telescope Narrow-band Imaging and
Spectroscopy\label{ttf}}

Our Anglo-Australian Telescope imaging was obtained with the
Taurus Tunable Filter \citep[TTF, ][]{bla98}, which is a Fabry-Perot etalon 
system.  Conditions were photometric
on the first night, and the typical image quality was $\sim 1.6$\arcsec. A
MIT-Lincoln Labs CCD was used behind the TTF.

The TTF was used with a full-width at half maximum 
(FWHM) spectral resolution of between 3.35 and 4.4\AA . Images were
obtained at central wavelengths of 4110\AA , 4114\AA , and 
4118\AA . 

An electronics fault was subsequently found to induce significant wavelength
drifts on timescales of a few hours. The 4110 and 4118\AA\ observations
were taken within four hours of a calibration, but the 4114\AA\ observations
were taken between four and eight hours after the calibration. Thus the 
wavelength calibration of all three images are somewhat uncertain, particularly 
that of the 4114\AA\ image. 

We obtained a long-slit spectrum of B1 with the Low Dispersion Survey 
Spectrograph \citep[LDSS, ][]{wyn88}. These observations are described
in \citet{F97}

\section{Results\label{results}}

\begin{figure*}

\epsscale{1.0}

\plotone{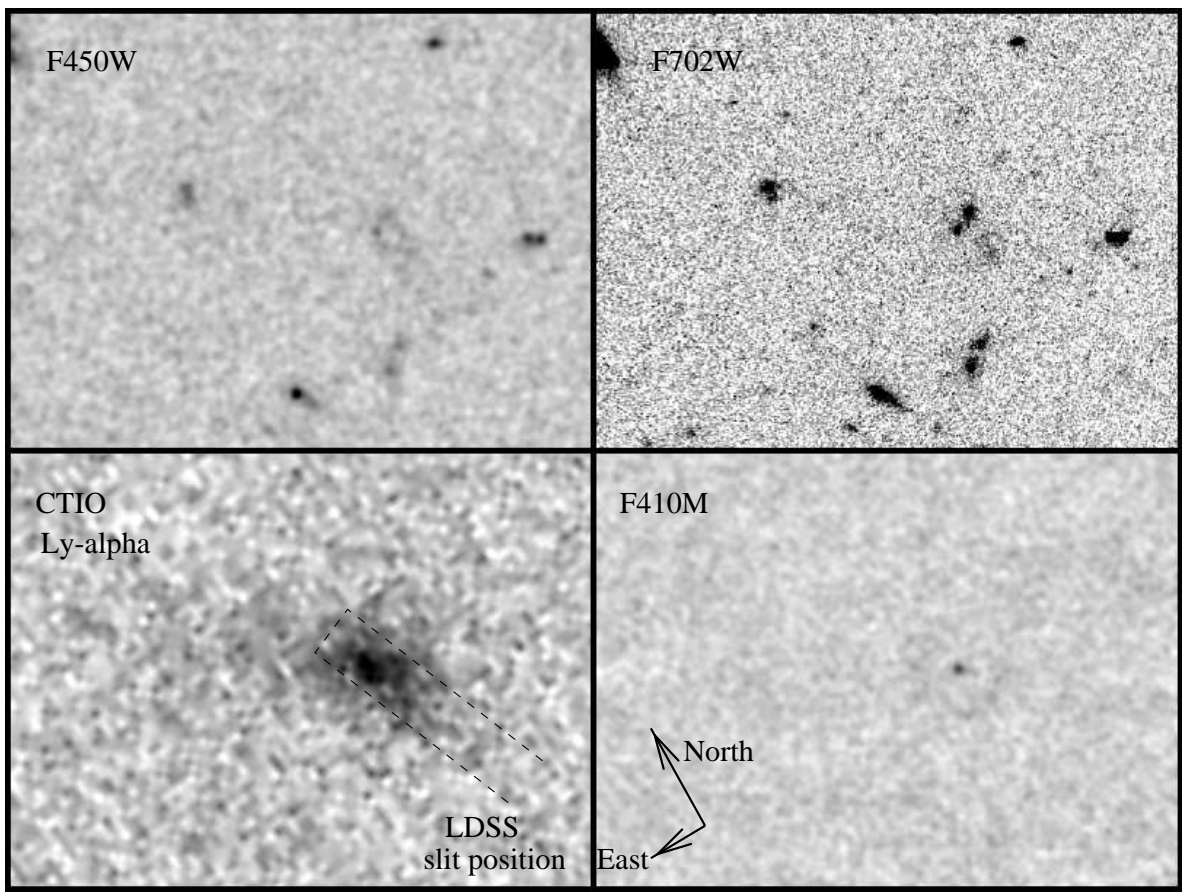}

\caption{Images of a 20\arcsec\ by 15\arcsec\ region
around the centre of B1. The bottom left image is the CTIO
Ly$\alpha$ image. The F410M and F450W images have been
smoothed with a Gaussian of $\sigma = 2$ pixels  (0.1\arcsec ). The CTIO 
image has been
rotated to match the orientation of the WFPC2 images. The continuum has been
removed from the Ly$\alpha$ images, and the Ly$\alpha$ component has been
removed from the F450W image.
\label{wfpc_frame}}

\end{figure*}

\begin{figure*}

\epsscale{1.0}

\plotone{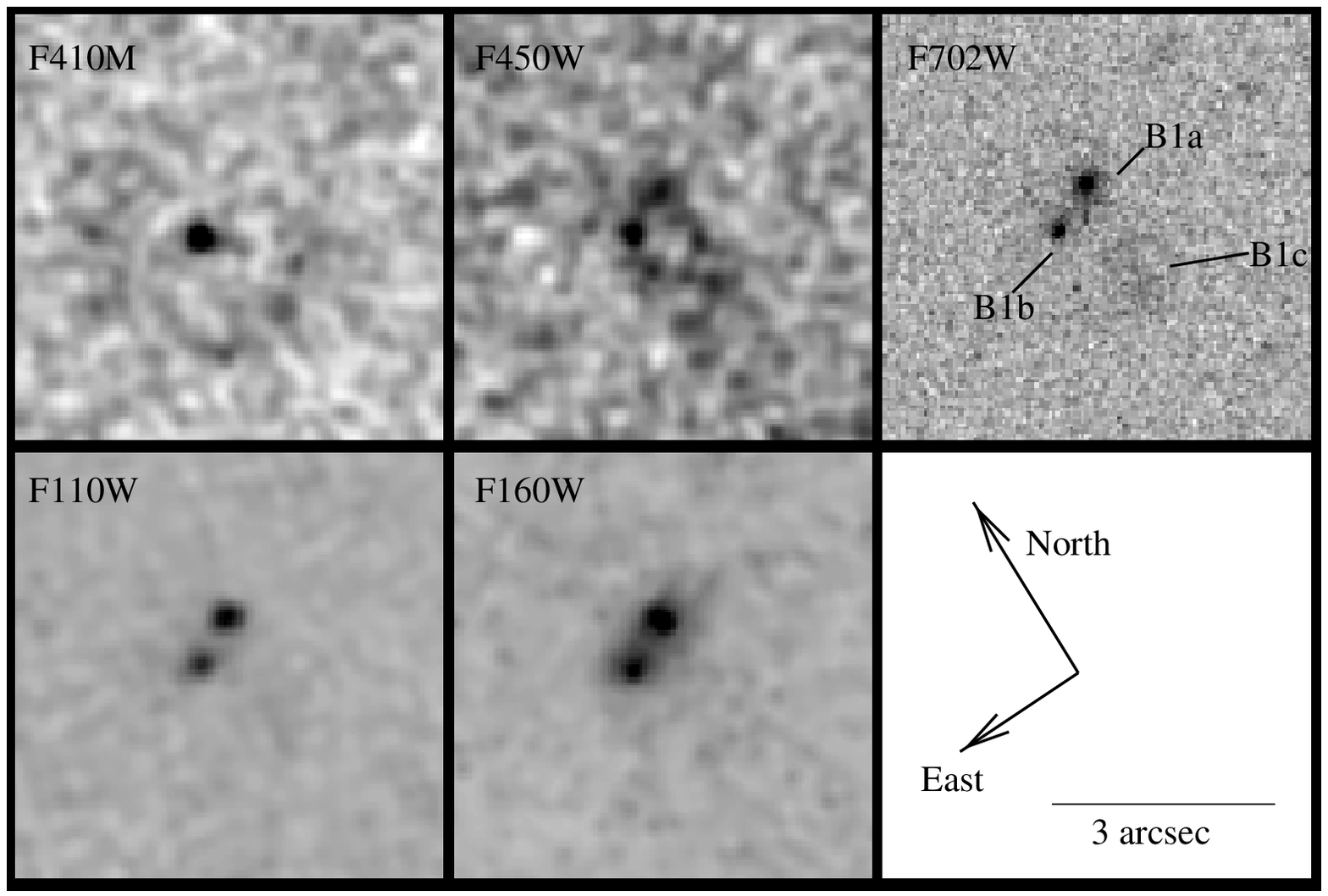}

\caption{Close-up of the continuum sources in
B1.  The F410M and F450W images have been
smoothed with a Gaussian of $\sigma = 2$ pixels (0.1\arcsec ). The NICMOS 
images
have been rotated to match the orientation of the WFPC2 images. 
The orientation matches that of Fig~\ref{wfpc_frame}. 
\label{zoom_frame}}

\end{figure*}

Our images of B1 are shown in Figures \ref{wfpc_frame} \& 
\ref{zoom_frame}. The appearance of B1 varies dramatically as a function of
wavelength and resolution. In the observed-frame red and near-IR, it breaks 
up into two compact components: B1a and B1b (Fig~\ref{zoom_frame}), 
separated by 0.8\arcsec . In this respect it is similar to another 
strong, high redshift Ly$\alpha$ source, the Coup-Foure\'e galaxy 
\citep{roc00}. A third continuum source, B1c, appears as chain 
of diffuse knots extending about 1\arcsec\ to the south from B1a and B1b.

In Ly$\alpha$, the picture is different again. When viewed with WFPC2,
we see a knot of Ly$\alpha$ emission from B1b, but no emission from B1a.
Diffuse patches of low surface brightness Ly$\alpha$ emission are found
$\sim 1$\arcsec\ to the south and north-east of B1b. The ground-based CTIO 
image has much greater sensitivity to low surface brightness
Ly$\alpha$ emission, and confirms that it extends at least 10\arcsec\ 
north-east of B1a and B1b, as reported by \citet{F96} and \citet{F97}. The 
non-detection of this emission in the WFPC2 image 
indicates that it is truly diffuse, and not coming from a series of compact
sources separated by less than the ground-based resolution limit (as was
the case for the Ly$\alpha$ nebula observed by \citet{pas96}).

The Ly$\alpha$ contribution to the blue continuum images, and the blue 
continuum contribution to the narrow-band Ly$\alpha$ images, were removed.
B1 has a total Ly$\alpha$ flux of about $1.5 \times 10^{-15}{\rm erg\ cm}^{-2}
{\rm s}^{-1}$: a more exact number is hard to calculate as the diffuse
flux fades gradually into the sky (this number is for a rectangular
aperture 10\arcsec\ long and 5\arcsec\ wide, aligned along the major axis of 
the Ly$\alpha$ flux). B1b has a Ly$\alpha$ flux of 
$\sim 2.0 \times 10^{-16}{\rm erg\ cm}^{-2}{\rm s}^{-1}$, which is only 
$\sim 13$\% of the total. The remaining flux is diffuse: the average 
Ly$\alpha$ surface brightness of the region within 2\arcsec\ of B1b is $\sim
4 \times 10^{-17}{\rm erg\ cm}^{-2}{\rm s}^{-1}{\rm arcsec}^{-2}$. 4\arcsec\
further east, the average surface brightness drops to $\sim 10^{-17}
{\rm erg\ cm}^{-2}{\rm s}^{-1}{\rm arcsec}^{-2}$.

\subsection{Fluxes}

Continuum fluxes (Table~\ref{magtab}) were calculated for the three
components of B1. Small circular apertures were used: 0.3\arcsec\ in
radius for B1a and B1b, and 0.65\arcsec\ for B1c. These small apertures
will miss much of the flux of these components, but were chosen to
avoid contamination from the other components. No aperture corrections
were attempted. A total flux for B1 was also 
measured, using a 2\arcsec radius aperture. 

Error estimates were made using apertures placed randomly on sky regions.
A copy of the image was made with all the sources detected
by the SExtractor \citep{ba96} package masked out. We randomly placed
1000 apertures of identical radius on the image. Any aperture that fell
on a masked region was rejected, and the procedure continued until 1000
clean measurements had been obtained. The standard deviation of these
measurements is our sky error estimate. A Poisson noise model was combined
with the sky error estimate to calculate object flux errors.
The Ly$\alpha$ contribution to the
F450W flux has been removed, but the probable emission-line contributions
to F110W, F160W and $K_s$ have not been removed: these contributions
are discussed in Section~\ref{linesub}. Finally, the fluxes were
corrected for the local Milky Way extinction ($E(B-V)=0.019$), as estimated 
by \citet{sch98}. The extinction curve of \citet{car89} was used.

Integrated near-IR magnitudes for B1 were measured from the NTT image, using
SExtractor with 4.2\arcsec\ radius circular apertures. We measure
$J=21.71\pm 0.12$, $H=20.31\pm 0.08$ and $K_s = 19.47 \pm 0.10$.

The measured coordinates of the components of B1 are shown in 
Table~\ref{coords}.

\begin{deluxetable}{lc}
\tablewidth{0pt}
\tablecolumns{4}
\tablecaption{B1 Component Coordinates \label{coords}}
\tablehead{
\colhead{Component} &
\colhead{Coordinates (J2000)} }
\startdata
B1a & 21:42:27.45$-$44:20:28.69 \\
B1b & 21:42:27.51$-$44:20:28.99 \\
B1c & 21:42:27.46$-$44:20:30.2  \\
\enddata
\end{deluxetable}

\begin{deluxetable}{lcccc}
\tablewidth{0pt}
\tablecolumns{5}
\tablecaption{B1 Component Observed Flux Densities \label{magtab}}
\tablehead{
\colhead{Filter} &
\multicolumn{4}{c}{Flux 
($ \times 10^{20}{\rm erg\ cm}^{-2}{\rm s}^{-1}{\rm \AA }^{-1} $) } \\
\colhead{ ~ }  &
\colhead{B1 (total)} &
\colhead{B1a} &
\colhead{B1b} &
\colhead{B1c} 
}
\startdata
F450W ($B$) & $70.2\pm 12.4 $ & $11.2\pm 1.3$ & $<7.26$\tablenotemark{a} & 
$25.7\pm 3.4$ \\
F702W ($R$) &  $64.2\pm 5.1 $ & $12.3\pm 0.8$ & $10.2 \pm 1.0$ & 
$15.3\pm 2.2$ \\
F110W ($J$) &  $47.8\pm 13.6 $ & $15.2\pm 0.8$ & $11.5\pm 0.8$ & 
$<8.35$\tablenotemark{a} \\
F160W ($H$) &  $120.2\pm 16.4 $ & $25.7\pm 0.9$ & $21.3\pm 0.9$ & 
$7.6\pm 1.2$ \\
\enddata
\tablenotetext{a}{3$\sigma$ upper limit}
\end{deluxetable}

\subsubsection{Emission-line Contamination\label{linesub}}

The near-IR filters are subject to possible emission-line
contamination at this redshift, z=2.38. [\ion{O}{2}] (3727~\AA ) is redshifted
into the $J$-band, H$\beta$ and the [\ion{O}{3}] (4959 and 5007~\AA ) doublet
are shifted into the $H$-band, and H$\alpha$ and [\ion{N}{2}] (6583~\AA ) are 
shifted into the $K$ band. These lines could in principle be very
strong \citep[eg.][]{eal93}. In this section, we estimate their
strength, and conclude that it is probably small.

As described in \citet{F96}, we imaged the field with a 
narrow-band filter of
wavelength coverage $2.238 \pm 0.024 \mu$m, which includes both
H$\alpha$ and [\ion{N}{2}]. Using a 5\arcsec\ radius aperture, we determine 
that 
these lines contribute $21 \pm 10$\% of the integrated $K_s$ flux of B1.
The F164N NICMOS image covers the wavelength of
redshifted H$\beta$. No significant narrow-band excess was detected: we
can place a $3 \sigma$ upper limit on the fraction of the integrated
$H$-band light coming from H$\beta$ in B1 of 3\% .

We have no direct measurements of the other line fluxes, so we estimate
their importance by making assumptions about their ratios with measured 
Balmer lines. Typical line ratios were taken from \citet{om93},
\citet{ken92} and \citet{tep00}, for a variety of assumed chemical 
compositions and ionization sources, and including both narrow- and
broad-line AGN ratios.

We conclude that line emission probably contributes only $\sim 20$\% at
most of the total continuum flux of B1 in $H$ and $K_s$. In $J$, the
contribution is probably $\sim 5$\% . Furthermore, only 
about 10\%~of the line flux will likely come from within the
photometric aperture of B1b, and $\sim 20$\%~from within the aperture
of the B1c.  The remainder of the optical emission lines will come
from the region of the diffuse blob. As B1a shows no Ly$\alpha$ emission,
its rest-frame optical line flux is also probably small.

The line corrections will thus probably be minimal. We have not, therefore,
applied any emission-line corrections to the near-IR fluxes.

\subsection{Velocity Structure\label{velocity}}

The Ly$\alpha$ line is generally extremely optically thick and often
strongly self absorbed, making it a very unreliable tracer of dynamics.
It is, however, the only sufficiently strong line available to us, and can 
be used to place some limits on the dynamics of the Ly$\alpha$ nebula 
surrounding B1.

\begin{figure*}

\epsscale{1.0}

\plotone{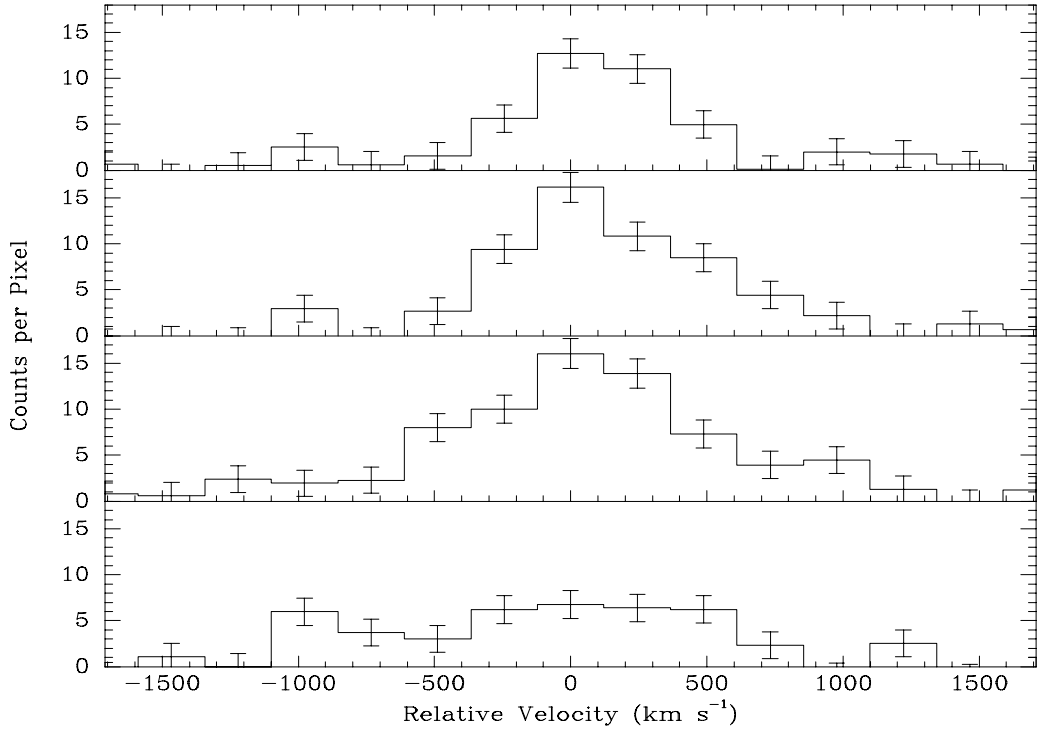}

\caption{Longslit Ly$\alpha$ spectra of B1. The slit orientation is shown
in Fig~\ref{wfpc_frame}. The four panels are the spectra 
extracted from the four adjacent CCD columns along which Ly$\alpha$ was 
detected: the top panel is the northernmost. Each pixel is 0.83\arcsec\ 
wide: the seeing was 1.2\arcsec , so the spectra are not independent.
Positive velocities represent redshifts. Spectral resolution is
$700 {\rm km\ s}^{-1}$, so the line core is not resolved in the top
three panels. The velocity zero point is arbitrary.
\label{longslit}}

\end{figure*}

The velocity dispersion of the gas around B1 (measured at the peak of
the surface brightness, as seen in ground-based images) is 
$\sim 600 {\rm km\ s}^{-1}$ \citep{F96}. Ly$\alpha$ is redshifted
with respect to \ion{C}{4}, indicating probable self absorption. The 
deep multislit spectroscopy
of \citet{F97} marginally spatially resolves this emission along the
slitlet. Fig~\ref{longslit} shows the spectra extracted from different
CCD columns: notice the extra component at $+1000{\rm km\ s}^{-1}$ in
the southernmost spectrum (and another possible component at
$-1000{\rm km\ s}^{-1}$).

The diffuse Ly$\alpha$ structure differs significantly between the three
TTF images.  This confirms the existence of multiple
velocity components separated by $\sim 700 {\rm km\ s}^{-1}$. It is hard to
say more, given the uncertainty in the wavelength calibration of the
TTF data (Section~\ref{ttf}).

The STIS spectrum detected the Ly$\alpha$ flux from B1b. A flux of
$1.57 \pm 0.40 \times 10^{-16}{\rm erg\ cm}^{-2}{\rm s}^{-1}$ was
measured, consistent with the value derived from the WFPC2 imaging.
The line was not significantly spectrally resolved; an upper limit on the
velocity dispersion of $\sim 1400 {\rm km\ s}^{-1}$ can be placed.

\section{The Continuum\label{continuum}}

\subsection{The Colors\label{sed}}

\begin{figure*}

\epsscale{1.0}

\plotone{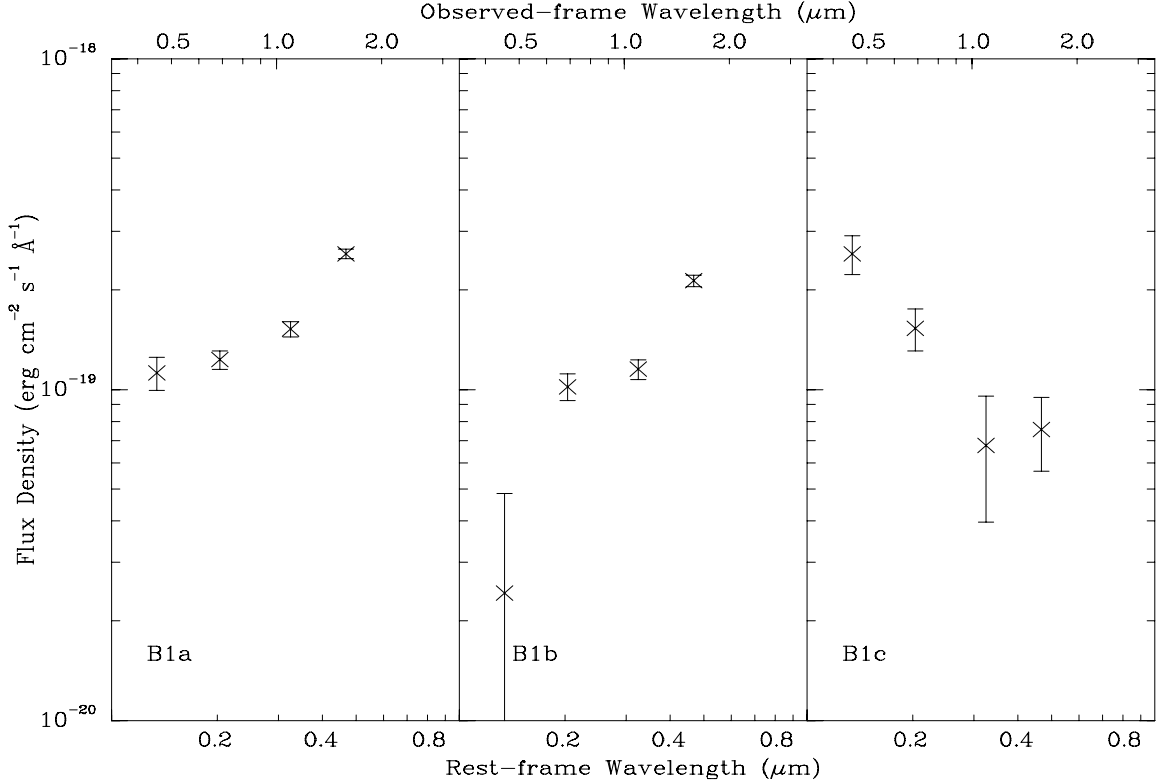}

\caption{The spectral energy distributions of the three components
of B1. Ly$\alpha$ emission has been subtracted
from the bluest point. \label{seds}}

\end{figure*}

The spectral energy distributions (SEDs) of the three components of
B1 are shown in Fig~\ref{seds}. B1a and B1b are quite red, while 
the SED of B1c is very blue. B1c's colors are similar
to those of Lyman-break galaxies \citep[eg.][]{ste96}. A power-law
fit to its colors gives a slope of the form $F_{\lambda} \propto
\lambda^{-1.34}$, which is close to the median slope of Lyman break
galaxies \citep[1.5, ][]{ade00}.

Our NTT photometry demonstrates that while B1 is red at wavelengths
shortward of $H$, it has a relatively blue $H-K_s$ color. This confirms
the result of  \citet{F96}.  B1's SED thus peaks in the $H$-band. As
we discussed in Section~\ref{linesub}, this is unlikely to be an
artifact of emission-line contamination.

\subsubsection{Modelling\label{sedmodel}}

\begin{figure*}

\epsscale{1.0}

\plotone{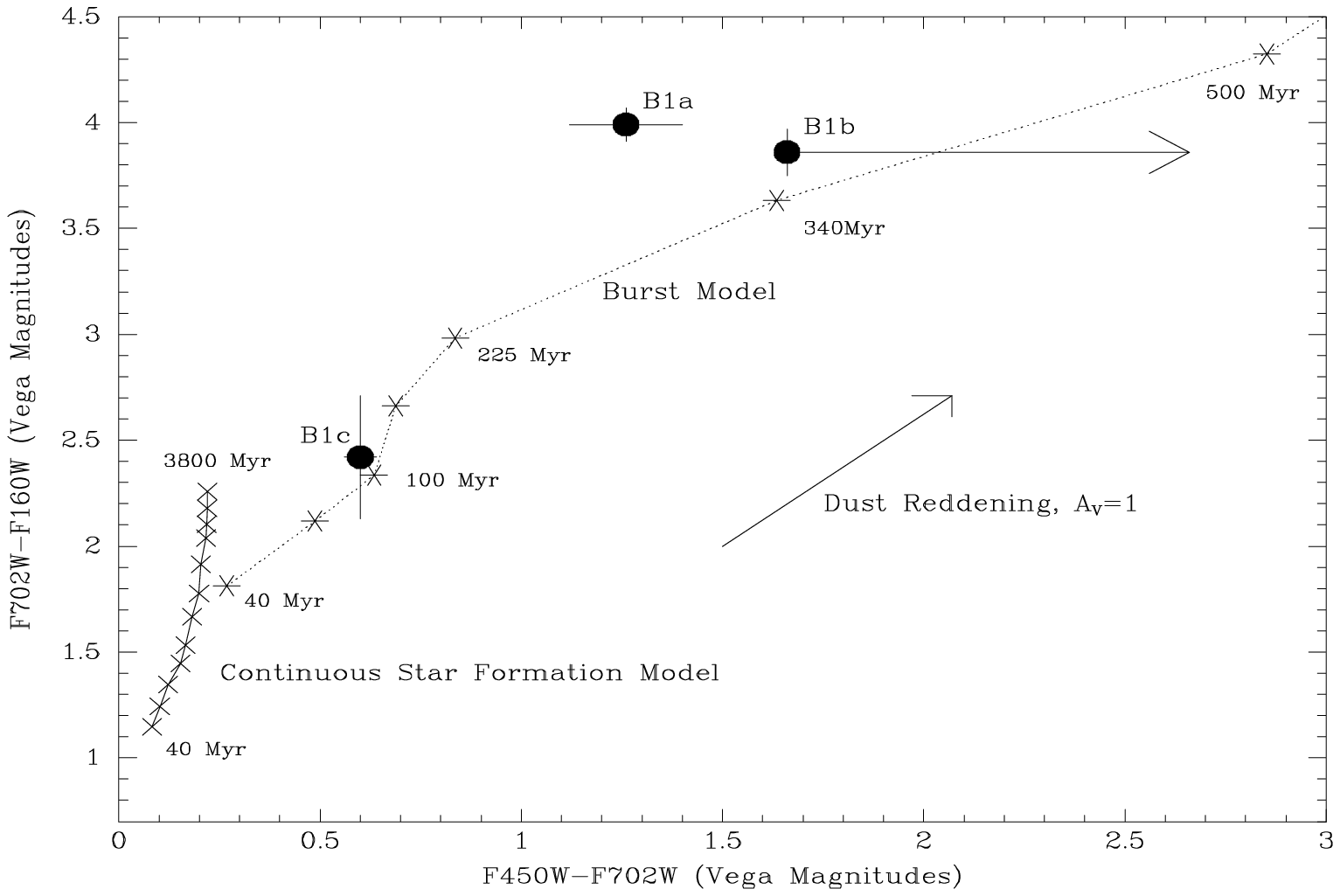}

\caption{The F450W$-$F702W and F702W$-$F160W colors of the three
components of B1. Both the continuous star formation models (crosses) and the
instantaneous burst models (asterisks) have been computed for twelve 
different ages of the stellar population. From left to right,
the model ages are 40, 60, 100, 150, 225, 340, 500, 750, 1125, 1700,
2500 and 3800 Myr. The change in colors caused by
dust with a rest-frame $V$-band extinction $A_v$ of one magnitude
is shown.\label{brhmodel}}

\end{figure*}

\begin{figure*}

\epsscale{1.0}

\plotone{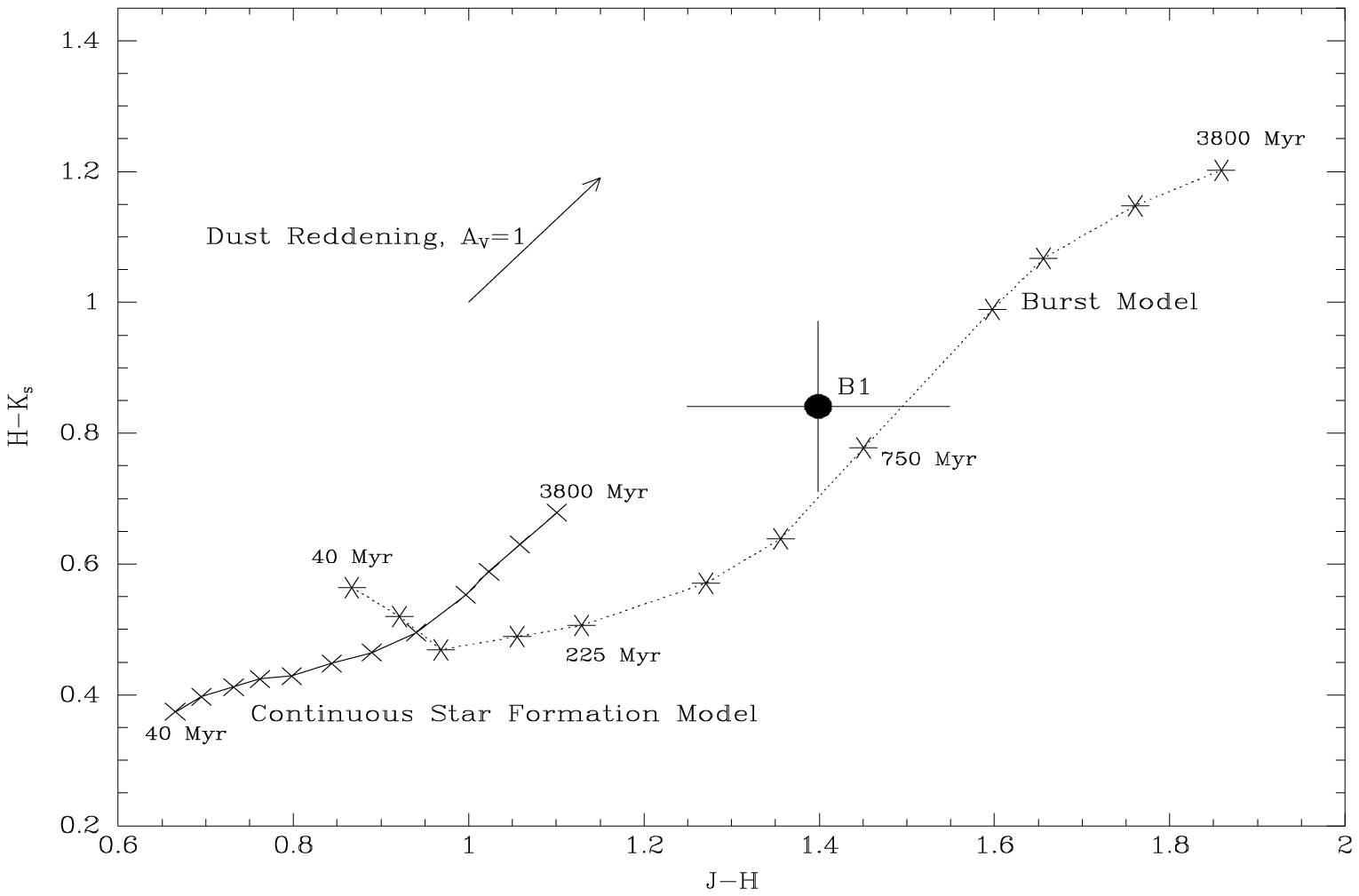}

\caption{The integrated NTT $J-H$ and $H-K_s$ colors of B1. Model
ages and extinction arrow as in Fig~\ref{brhmodel}.\label{jhkmodel}}

\end{figure*}

We modelled the SEDs of B1 using the 1997 version of the spectral synthesis 
models of \citet{bru93}. Models both with continuous uniform star formation 
and models in which all the stars were formed in an instantaneous burst (a 
simple stellar population) were used. All models had Salpeter stellar initial 
mass functions, with no mass cut-off. Dust was modelled using the empirical 
absorption curves of
\citet{cal94}. Our predicted colors are compared with the observations in
Figures \ref{brhmodel} and \ref{jhkmodel}.

The red components, B1a and B1b, can be fit by two models: 

\begin{itemize}

\item An unreddened burst model of age $750\pm 150$ Myr (1$\sigma$ limits)
and stellar mass $\sim 8 \times 10^{10} M_{\Sun}$. This naturally fits the 
$H$-band peak
(Fig~\ref{jhkmodel}) and the photometry from $F702W$ thru $K_s$. It
greatly underpredicts the observed F450W flux of B1a. We do not consider this
to be a problem: a star formation rate of only $\sim 0.1 M_{\Sun}{\rm
yr}^{-1}$ will produce OB stars sufficient to produce the observed blue
flux.

\item A dusty starburst model. The best fit is a model with a continuous
star formation rate of $\sim 10^3 M_{\Sun} {\rm yr}^{-1}$, an extinction
of $A_v = 2.3$ mag, and an age of 500 Myr. A slightly worse fit (but still 
acceptable at the $2\sigma$ level) is younger (100 Myr), dustier ($A_V =3$
mag) and has a star formation rate of $\sim 10^4 M_{\Sun}{\rm yr}^{-1}$.

\end{itemize}

The dusty starburst model is hard to reconcile with the strong Ly$\alpha$
emission of B1b, as Ly$\alpha$ is resonantly scattered and thus very strongly
absorbed by even small amounts of dust.

B1c is much bluer, and can be fit either by an unreddened instantaneous burst
model of age $\sim 100$ Myr and mass $\sim 3 \times 10^9 M_{\Sun}$, or by
a slightly dusty continuous star formation model, with a star formation
rate of $\sim 10 M_{\Sun}{\rm yr}^{-1}$ and $A_V \sim 1.0$.

\subsection{The Radial Profiles\label{psf}}

In this section, we show that B1a and B1b are spatially extended.
They are, however, very compact objects: too compact to be easily
modelled as spiral galaxy disks. They appear to show colour gradients:
bluer in their central regions.

\begin{figure*}

\epsscale{1.0}

\plotone{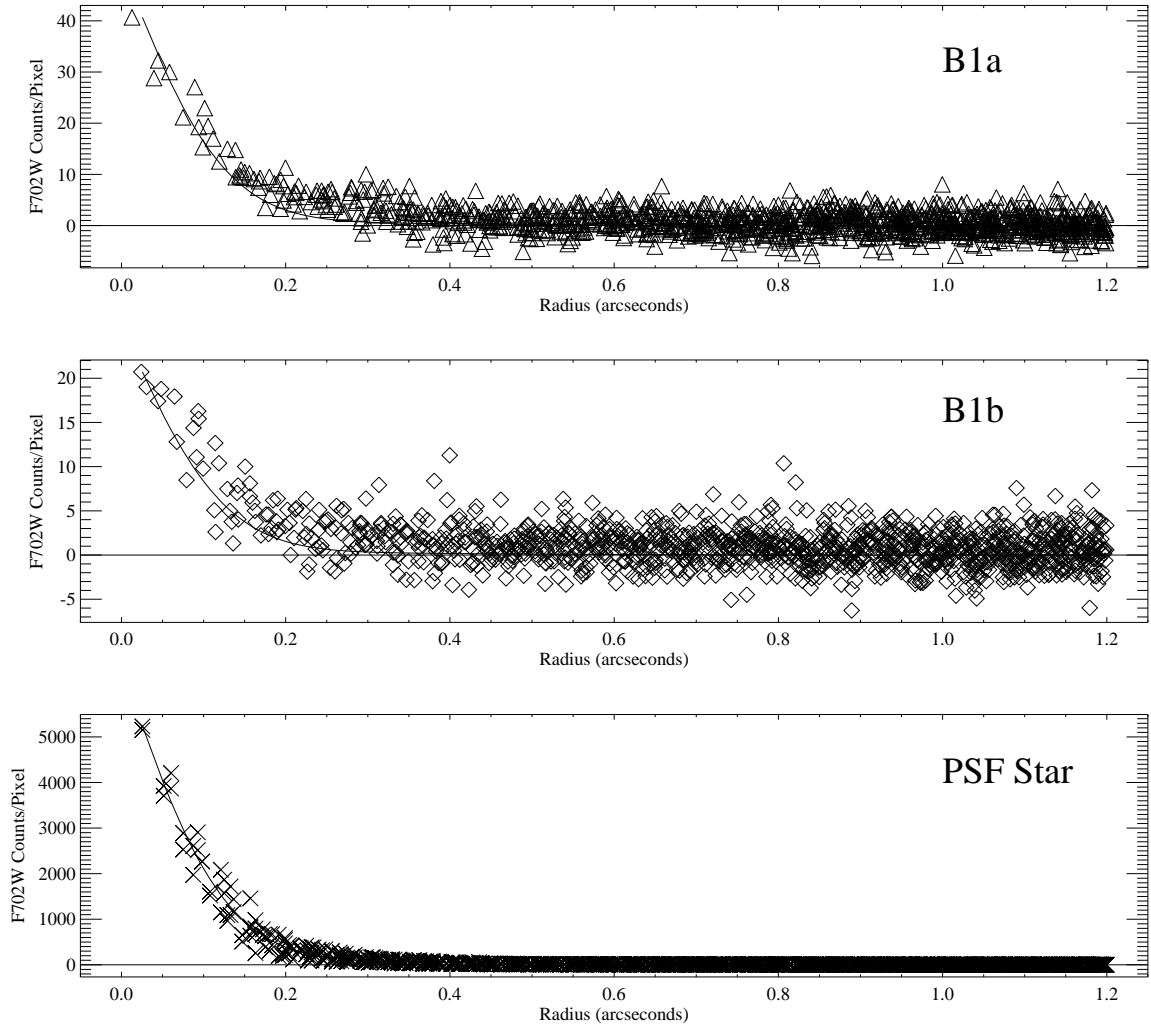}

\caption{The F702W radial profiles of B1a, B1b and a point source
image on the same chip. 
The solid line is an analytic fit to the PSF and is the same in
all panels. The quadrant of each component that faces the other component
has been excluded from the plots.
\label{rpsf}}

\end{figure*}

\begin{figure*}

\epsscale{1.0}

\plotone{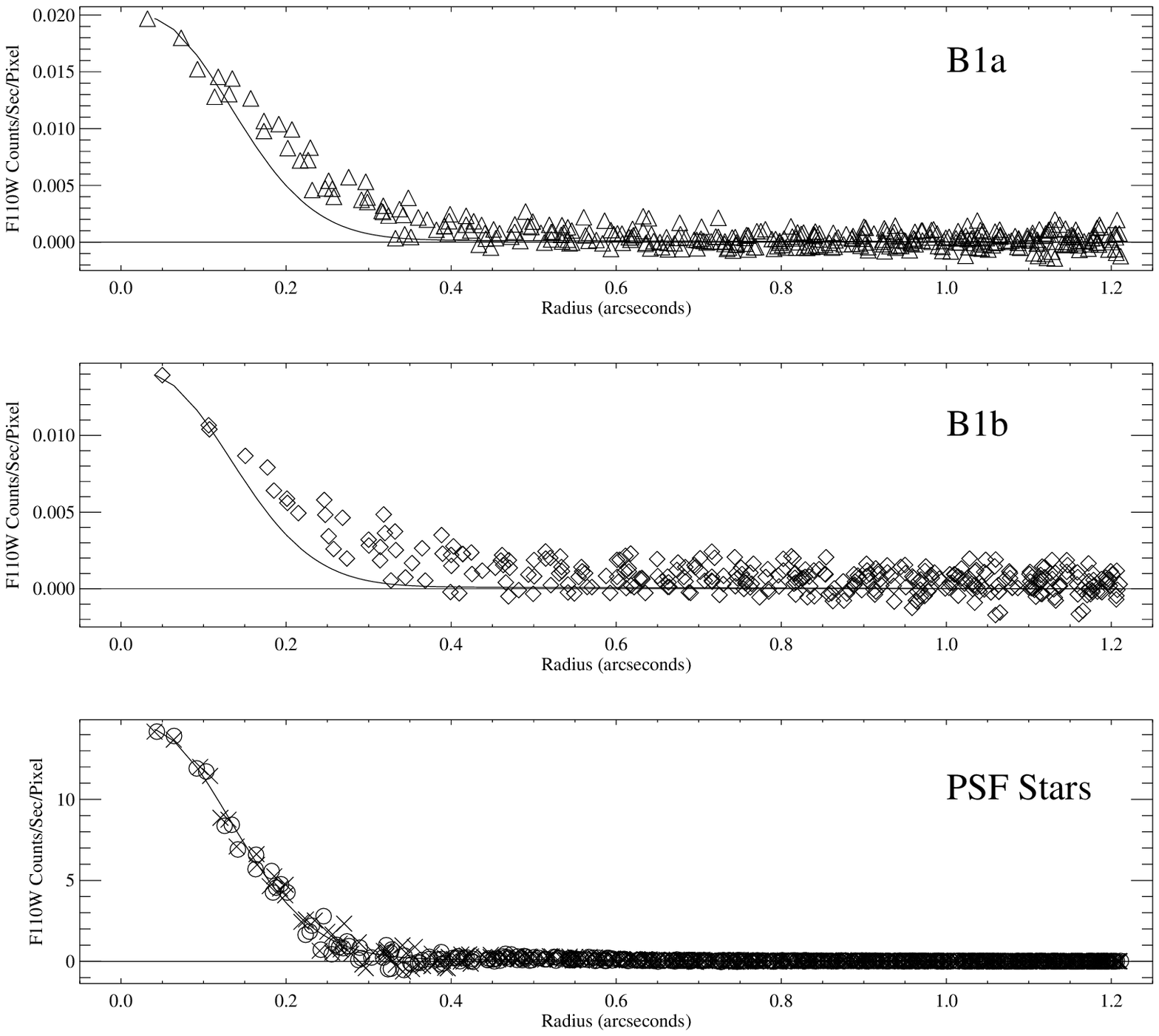}

\caption{The F110W radial profiles of B1a, B1b and a point source from
the same image. 
The solid line is an analytic fit to the PSF and is the same in
all panels. The quadrant of each component that faces the other component
has been excluded from the plots. \label{jpsf}}

\end{figure*}

In Figures~\ref{rpsf} and \ref{jpsf} we compare the radial profiles
of B1a and B1b with the relevant point spread functions (PSFs) measured
from a bright star in the same image.
B1a and B1b are marginally resolved in the
F702W (rest-frame 2100 \AA ) and F110W (rest-frame 3250 \AA ) images. 
The F160W (rest-frame 4730\AA ) image was taken with NICMOS out of
focus, and it does not significantly resolve B1a and B1b.
The observed PSF does not vary significantly across the images, and
is consistent with analytic predictions. 

\begin{figure*}

\epsscale{1.0}

\plotone{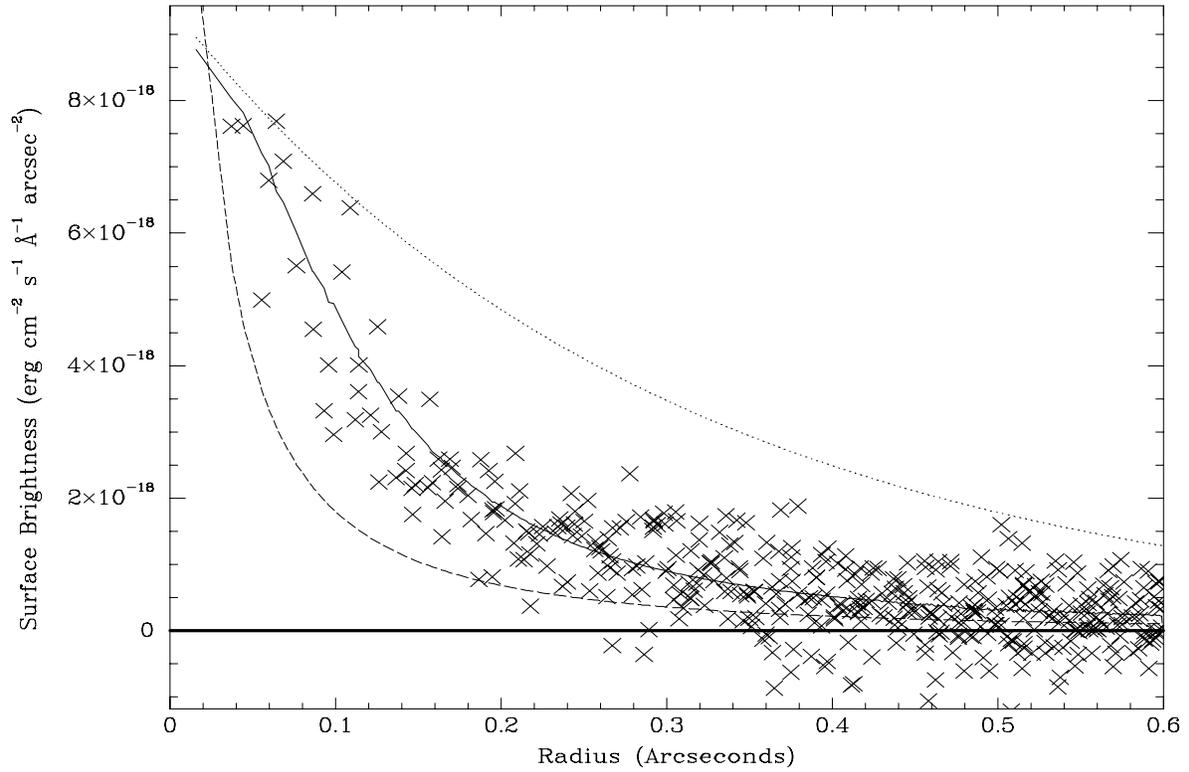}

\caption{The F702W radial profile of B1a. The dashed curve (left) is
a model spheroid with $R_e = 0.5\arcsec $. The dotted line (right) is
a model disk with $R_d = 0.3\arcsec $. Both radii are typical of redshift
zero galaxies. Neither model has been
convolved with the PSF. The solid line is a spheroid with $R_e = 0.5\arcsec $, 
convolved with the PSF and scaled to fit.
\label{newfig}}

\end{figure*}

Although B1a and B1b are clearly resolved, both are very compact, with
the observed surface brightness dropping to half its peak value
within about 0.1\arcsec\ ($\sim 1$ kpc). This places a strong constraint
on the size of the galaxies. Consider a disk galaxy, with a typical
radial surface brightness profile of the form
\[
I(r) = I_d  \exp(-R/R_d)
\] \citep{bin98}.
The surface brightness of such a disk drops to half its peak value at 
$r_{1/2} = 0.69 R_d$ (Fig~\ref{newfig}). Thus our observations require that 
$R_d < 0.15\arcsec\ \sim 1.2$ kpc. Only $\sim 2$\%
of modern galaxy disks are this compact \citep{ken84}. Now consider a 
spheroidal galaxy, with a radial surface brightness profile of the form
\[
I(r) = I_e \exp\{-7.67[(R/R_e)^{1/4}-1]\}
\] \citep{bin98}.
This profile is much more sharply peaked than the exponential disk
profile, and its surface brightness drops to half its peak
value at $r_{1/2} = 0.13 R_e$ (Fig~\ref{newfig}). 
Thus $R_e < 0.75\arcsec\ \sim 6$ kpc.
Over 50\% of modern elliptical galaxies have effective radii this small
\citep{djo87}.

More detailed modelling confirms these results. A wide variety of
model galaxy profiles were convolved with PSFs and fit by
$\chi^2$ minimisation to the observed radial profiles.
The modelling confirms that B1a and B1b are significantly extended.
Both exponential disk models and spheroidal (de Vaucouleurs) models
give acceptable fits to the data, as do models with a point source
embedded in a fainter halo. Pure exponential disk models only give
acceptable fits if $R_d < 0.1\arcsec\ $ ($\sim 800$  pc), which would place
B1a and B1b in the most compact 1\% of present-day disk galaxies. Spheroidal 
models give best fits for effective (half-light)
radii $R_e$ of around $0.5$\arcsec\ ($\sim 4$ kpc), which
are typical of modern elliptical galaxies. 
Models which combine a nuclear spheroid or point source with a disk
can give acceptable fits, but only if the disk is unusually small and/or
contributes only a small fraction ($<$ 10\%) of the light.

The compact radial profiles of B1a and B1b are typical of those of
high redshift Lyman-break galaxies \citep{gia96}. The inferred
surface brightnesses in the rest-frame UV are comparable to those of
faint Lyman-break galaxies, but the rest-frame optical surface 
brightnesses are an order of magnitude higher.  As \citet{gia96} note,
these compact radial profiles are consistent with present-day elliptical
galaxy profiles or bulges, but are much more compact than present-day disks.

Curiously, both B1a and B1b appear to be significantly larger at longer 
wavelengths. This can be seen in Figures~\ref{rpsf} and \ref{jpsf}:
despite the wider PSF for the F110W filter, both components are
clearly more extended than as seen through the F702W filter. Our
modelling confirms this result:
even after convolution with the relevant PSFs, no single model gives
a good fit to either B1a or B1b at both F702W and F110W. The central
$\sim 1$ kpc of both components is $\sim 0.3$ magnitudes bluer in
F702W$-$F110W than the outer regions.

The change in size as a function of wavelength could be explained by a colour 
gradient within a single spheroidal component. It could also be explained 
by a two component model: a compact blue component that dominates at
rest-frame 2100\AA\ and a more diffuse (but still compact) red component that
dominates at 3250\AA\ and beyond. This red component, if disk-like,
must have $R_s < 0.2\arcsec $, which would still place it in the most compact
1\% of low redshift disks.

\subsection{Discussion}

What are the continuum components of B1? B1c, with its blue colors, 
appears relatively 
straightforward: it is a region of extended moderate star formation,
obscured by little dust. This star formation may be triggered by the
interaction of B1a and B1b.

The nature of B1a and B1b is less straightforward. We consider four
models in turn.

\subsubsection{Elliptical Galaxies}

Could B1a and B1b be young elliptical galaxies? This hypothesis fits
the data well.

\begin{itemize}

\item  The radial profiles of B1a and B1b are well fit with de Vaucouleurs
profiles, and the inferred radii are typical of present-day elliptical
galaxies.

\item  The red colors are most easily explained by a $> 10^{11} M_{\Sun}$ 
stellar population which completed its star formation about 750 Myr before we 
observe it. It would thus be similar to modern E$+$A galaxies.

\item  B1 lies in an overdense region of the early universe \citep{F00},
which may be the ancestor of a galaxy cluster.

\item The velocity field around B1, if virial, implies masses of $\sim 10^{12}
M_{\sun}$.

\end{itemize}

If B1a and B1b are elliptical galaxies, how do their surface brightnesses
and sizes fit on the Kormendy relation \citep{kor77,hoe87}? The F160W filter
corresponds quite closely to the rest-frame $B$-band. The radial profile
fitting suggests that both components have effective radii in the
observed-frame near-IR of $\sim 4$kpc. Integrating a de Vaucouleurs
profile over our photometric aperture, we can convert our observed $F160W$ 
magnitudes into
rest-frame $B$-band surface brightnesses at the effective radius. Were
these components at low redshift, their surface brightnesses at the effective
radius would be $B_{0V} \sim 19.6{\rm \  mag\ arcsec}^{-2}$. This is roughly
an order of magnitude higher than the Kormendy relation would predict
for an elliptical galaxy of this radius. This is consistent with
passive evolution from our unreddened burst model (Section~\ref{sed}).

If B1a and B1b are elliptical galaxies, why do they apparently show 
color gradients? The UV emission from the central $\sim 1$ kpc could be
caused by a starburst, a hole in the dust or (for B1b) an AGN. Alternatively, 
some chemodynamical models predict color gradients in young elliptical 
galaxies \citep[eg.][]{fri98,jim99}.

\subsubsection{Dusty AGN}

A dusty AGN can certainly have colors as red as B1a and B1b 
\citep[eg.][]{FWW00}. The red components of B1a and B1b are however
spatially extended, and hence cannot be produced by an AGN. Could
the observed color gradient, however, be caused by the superposition of
a compact blue AGN on a more extended, lower surface brightness red galaxy? 

Even in the relatively blue central regions, the continuum slope is
redder than $F_{\nu} \propto \nu^{-2}$, while radio-quiet QSOs have
typical continuum slopes of $F_{\nu} \propto \nu^{-0.5}$ \citep{FWW00}.
Any central AGN would thus have to be reddened by dust with $E(B-V) >
0.2$. This would make it hard for the Ly$\alpha$ emission from B1b
to escape. In addition, the AGN in B1a and B1b would need remarkably equal 
luminosities and dust extinctions to explain the very similar observed 
colors and
radial profiles of the two components. This fine tuning seems implausible.
Furthermore, the high measured surface brightness was
measured in the F160W band, to which the central blue component does not
significantly contribute. We therefore conclude that while AGN may be 
present, they are unlikely to be responsible for the observed colors.
Sensitive hard X-ray or far-IR observations will be required to determine 
whether a dusty AGN is present.

\subsubsection{Dusty Disk Galaxies}

The more extended red component of B1a and B1b can be fit by a very
compact exponential disk. Could these two objects 
then be compact spiral
galaxies? The more compact bluer central component could be caused
by AGNs or nuclear starbursts.

With sufficient dust, a spiral galaxy can certainly appear as red as
these two components (Section~\ref{sedmodel}). Two lines of reasoning 
however oppose this hypothesis.

\begin{enumerate}
 
\item Even the extended
red components are extremely compact by the standards of modern disk
galaxies. This could, however, be explained if disk galaxies form
from the inside outwards.

\item This model requires that the outer (red)
components of B1a and B1b be greatly reddened. This would make it difficult
for the observed UV light and Ly$\alpha$ emission from the bluer central 
components to escape.

\end{enumerate}

\subsubsection{Bulges plus Low Surface Brightness Disks}

Could B1a and B1b be the bulges of two disk galaxies? The
disks themselves may not have formed, or might be too low in surface
brightness to be detected by the Hubble Space Telescope.

Two lines of reasoning oppose this hypothesis. Firstly, the red colors
and $H$-band peak of B1 are seen in our ground-based photometry with large
($\sim 5\arcsec$) apertures. Thus any extended component
must have a similar stellar population to that inferred here. Secondly,
the inferred masses and sizes for B1a and B1b are far larger than those
typical of the bulges of disk galaxies.

\section{The Ly$\alpha$ Nebula\label{lines}}

B1 has a total Ly$\alpha$ luminosity of $\sim  
 10^{44}{\rm erg\ s}^{-1}$, spread over a region at least
$30 \times 100$ kpc in size. The emission consists of a number of
diffuse, discrete components
with relative velocities of $\sim 700 {\rm km\ s}^{-1}$.
What can produce such a luminous, diffuse, fast moving 
Ly$\alpha$ nebula? 

The bulk motions of the Ly$\alpha$ nebula suggest that fast shocks must
be present. We will show that such shocks are quite capable of producing the
observed Ly$\alpha$ luminosity. The puzzle then becomes one of explaining the
origins of the bulk motions.

It has also been suggested that photoionzation by an AGN, or
a cooling flow may power such nebulae. We discuss these possibilities, which
are hard to exclude.

\subsection{Physical Parameters of the Nebula}

If we assume that the Ly$\alpha$ velocities are representative of the
gas, and not an artifact of the high optical depth in Ly$\alpha$, then 
the dynamical timescale (crossing time) of the nebula is $\sim 10^8$ 
years. This
is interestingly similar to the inferred age of the stellar population 
(Section~\ref{sed}).

What is the density of the nebula? The sight-line to background 
QSO 2139$-$4434 passes 20\arcsec\ from B1 (though it does not pass through 
the Ly$\alpha$ nebula). The QSO spectrum shows a strong absorption-line 
system at the redshift of B1, with a column density 
$N_H \sim 10^{19}$ \citep{F00}. If we assume that the column density through 
the Ly$\alpha$ nebula is at least as great as this, and that the nebula
is $\sim 20$ kpc thick (ie. as thick as it is wide) along the line 
of sight (both big assumptions), 
then its density must be at least $10^{-4} {\rm cm}^{-3}$. 
If, however, the density were this low, the gas 
would be fully ionized by the metagalactic UV background at this
redshift. If the nebula is $\sim 20$ kpc thick, it must have a density of
$> 10^{-2}$ for its recombination rate to balance the photoionization
from the UV background \citep{F00}. The diffuse star formation of B1c, 
taking place within part of the Ly$\alpha$ nebula, also suggests that
densities may be higher. In reality, of course, the cloud may be
highly inhomogeneous, with dense neutral `bullets' embedded in low
density ionized gas.

If we take this lower limit on the density of the nebula ($N_H > 10^{-2}
{\rm cm}^{-3}$), and assume that the nebula is 100 kpc long, 30 kpc
wide and 10 kpc thick, then its hydrogen mass is $\sim 10^{10} M_{\sun}$.

\subsection{Shock Models\label{shock}}

Whenever there are blobs of gas moving at supersonic relative speeds, as
are observed in the nebula, shocks are inevitable. We thus know that shocks
are present. Could the emission from these shocks produce the observed 
Ly$\alpha$ emission?

\subsubsection{Modelling}

We used the fast shock models computed with the MAPPINGS~III code to
calculate the line emission from the shocks. MAPPINGS~III is  
an updated version of the MAPPINGS~II code \citep{dop96} and  
uses improved radiation continuum and hydrogenic calculations to calculate
the line and continuum radiation from shocks.  The 
new version of the code enabled the calculation of a grid of shocks 
for a range of metallicities, up to 1000 km/s, although a small 
number of low metallicity models above $800 {\rm \ km\ s}^{-1}$ failed 
to converge after several global iterations.  A fiducial model with 
a velocity of $700 {\rm \ km\ s}^{-1}$ was used here, having achieved 
a converged solution over a range of metallicities. 

If the shocks are radiative, and if we assume, for example, a density of 
$1 {\rm \ cm}^{-3}$, then
the Ly$\alpha$ emissivity will be $0.13 {\rm \ erg\ cm}^{-2}{\rm s}^{-1}$, 
almost independent of the metallicity of the gas. A quarter of this comes from 
the shock
itself, the rest coming from photoionization of the assumed neutral gas in 
front of the shock (the precursor region). If the shock is face-on to our 
sight-line, this would correspond to an observed surface brightness of 
$1.6 \times 10^{-15}{\rm erg\ cm}^{-2}{\rm s}^{-1}{\rm arcsec}^{-2}$:
nearly two orders of magnitude greater than the observed Ly$\alpha$
surface brightness (Section~\ref{results}).  Much of this may be self 
absorbed, but if even 1\% escapes, it can explain our observations.
The predicted Ly$\alpha$/\ion{C}{4} ratio is also consistent with the 
observations \citep{F96}.

How does this vary with the assumed density of the gas? If the shocks are
radiative, the emissivity will scale roughly with the density. Thus even
with our lower limit on the density $\rho = 10^{-2}{\rm cm}^{-3}$
(Section~\ref{lines}), shocks can produce all the Ly$\alpha$ we observe.

Would these shocks be radiative? For our lower limit on the density,  
$700 {\rm km\ s}^{-1}$ shocks become radiative on a timescale
of $\sim 10^8$ years, which is comparable to the dynamical timescale.
Thus it seems likely that at least some fraction of the shocks will be
radiative.

Thus shocks are probably present, and  
can easily produce the Ly$\alpha$ emission we observe. A number of factors
could, however, suppress the Ly$\alpha$ emission from shocks. 

\begin{enumerate}

\item The high optical depth in Ly$\alpha$ may have caused us to overestimate
the velocity dispersion of the Ly$\alpha$ nebula. The emission from shocks is
a strong function of their velocity: $200 {\rm km\ s}^{-1}$ shocks emit
two orders of magnitude less Ly$\alpha$ per unit area than our fiducial model.

\item The density of the Ly$\alpha$ nebula may be close to our lower
limit of $\rho = 10^{-2}{\rm cm}^{-3}$, in which case the Ly$\alpha$ emission,
after the probable self absorption losses, could be an order of magnitude 
below the observed Ly$\alpha$ surface brightness. 

\item The Ly$\alpha$ nebula could consist of small dense `bullets' of
neutral gas, moving at high speeds through a hot, highly ionized, low
density medium. The bowshocks in the ionized medium would not be
radiative, while the reverse shocks within the `bullets' would be slow,
and hence have low surface brightnesses (as well as having a small area). 

\end{enumerate}

\subsubsection{The Energy Source\label{energy}}

If the shock model is correct, the bulk gas motions explain the 
Ly$\alpha$ luminosity. What then could be the energy source for the bulk gas 
motions? 

If we divide
the total kinetic energy of the nebula (assuming a mass of $\sim 10^{10}
M_{\sun}$) by the Ly$\alpha$ luminosity, we can derive an upper limit on 
the damping time, which is $\sim 10^7$ years.
This is less than the dynamical timescale and stellar age of B1, 
suggesting that continued energy input is required.
Note that the Ly$\alpha$ we observe may only be a small fraction
of the Ly$\alpha$ emitted, which in turn will only be a small fraction 
($\sim 1$\%) of the total energy dissipation in the shocks.  On the other hand,
the gas mass may be greater than we are assuming. 

If continued energy input is required, what can the source be? 
Perhaps the most natural energy source is the gravitational potential
energy of B1. If we assume that B1 has a total mass of $\sim 10^{12} 
M_{\sun}$, then the gravitational potential energy released in
its formation would be  $\sim 10^{61}$ ergs, which is more than
sufficient to power the Ly$\alpha$ luminosity for the dynamical timescale.
Such a mass for B1 would imply virial velocities in the Ly$\alpha$
nebula of $\sim 500 {\rm km\ s}^{-1}$, which are comparable to the bulk gas 
motions observed. This energy could be released by the merger of B1a and B1b,
or by a continuing mass infall rate of $\sim 10^3 M_{\sun}{\rm yr}^{-1}$. 
\citet{hai00} point out that chemically primordial gas at a temperature of
$\sim 10^4$ K has few cooling mechanisms other than Ly$\alpha$ emission.
Thus if most of the gas around B1 is primordial, a Ly$\alpha$
luminosity comparable to the gravitational binding energy
(Section~\ref{energy}) of the system must be radiated.

Alternatively, 
\citet{tan00} suggest that galactic superwinds driven by a starburst
could power extended Ly$\alpha$ nebulae. This would naturally explain the
similarity between the dynamical timescale and the inferred age of the stellar
population in B1a and B1b. The maximum plausible amount of energy which
star formation can inject into the intergalactic medium is
$\sim 10^{49} {\rm erg\ M}_{\sun}^{-1}$ \citep[eg.][]{bow00}. As B1a and
B1b have a combined stellar mass of $\sim 10^{11} M_{\sun}$, the total energy
liberated would be $\sim 10^{60}$ ergs, which could 
in principle drive the observed Ly$\alpha$ luminosity
for $\sim 10^{10}$ years.

The Ly$\alpha$ emission in high redshift radio galaxies has sometimes
been ascribed to mechanical energy deposition by a radio jet
\citep[eg.][]{bic00}. Could such a jet be present in B1?
We do not detect any radio emission from B1 \citep{F96}. Our $3 \sigma$
upper limit of 0.23 mJy at 2.4 GHz corresponds to a luminosity
limit of $10^{33}{\rm ergs\ s}^{-1}{\rm Hz}^{-1}$ at rest-frame 1.4 GHz 
(assuming a typical radio galaxy radio spectrum of the form 
$F_{\nu} \propto \nu^{-0.8}$). Typical ratios 
of the total jet mechanical power to the monochromatic radio flux at 1.4 GHz
are $\sim 10^{11}$---$10^{12}  {\rm Hz}^{-1}$ \citep{bic98}. Thus we can
place an upper limit of $\sim 10^{45}{\rm erg\ s}^{-1}$ on the energy
injection from a radio jet. This is sufficient to power the observed
Ly$\alpha$ luminosity only if $> 10$\% of the mechanical energy is
converted into observable Ly$\alpha$ emission, which seems implausible.
Deeper radio observations should test this hypothesis.

\subsection{Other Models}

If the shock emission is suppressed, is there an alternative mechanism 
capable of producing the Ly$\alpha$ emission?

Stellar ionisation (by hot young stars) is incapable of producing the
observed \ion{C}{4} emission. Photoionisation by the UV continuum
radiation from a QSO can, however, produce both the Ly$\alpha$ and \ion{C}{4},
as discussed by \citet{F97}. We do not see strong UV continuum
emission from any component of B1, so such an AGN would have to be
concealed from our sight-line. \citet{cha00} detected strong sub-mm
emission from a Ly$\alpha$ blob at $z=3.09$, suggesting that a dusty
AGN or starburst was present, but we have no evidence for such a source
in B1.

Extended emission-line nebulosities are associated with many low
redshift cooling flow clusters, such as NGC 1275. As all high-redshift
Ly$\alpha$ blobs seem to lie in cluster environments, could the Ly$\alpha$
luminosity be associated with a cooling flow \citep[eg.][]{fab86}?

\citet{ste00} suggest that Ly$\alpha$ nebulae are caused by gas cooling 
from higher temperatures, in such a 
cooling flow. If each cooling hydrogen atom produces one Ly$\alpha$
photon, however, we require that $\sim 10^6 M_{\sun}{\rm yr}^{-1}$
of hydrogen be cooling onto B1. This would build the observed baryonic
mass of B1 in $< 10^5$ years. This is far
smaller than either the dynamical time or the age of the stellar
population in B1a and B1b, so the simple cooling model
seems implausible.

Note that the nebular emission-line luminosities of low redshift cooling-flow 
clusters are also orders of magnitude 
too great to be explained by this simple recombination mechanism 
\citep[eg.][]{voi97}. The cause of this discrepancy is still 
controversial even at low redshifts.

On purely energetic grounds, 
could the Ly$\alpha$ luminosity be powered by the thermal energy in the 
nebula? If we divide the thermal energy of $10^{10} M_{\sun}$ of 
hydrogen at $10^6$ K by the observed Ly$\alpha$ luminosity, we get a 
timescale of $\sim 10^7$ years: shorter than the dynamical timescale or
the stellar age. $10^{10} M_{\sun}$ may however be
an underestimate of the true gaseous mass of the Ly$\alpha$ nebula,
particularly if the neutral gas we observe is immersed in a much more massive
cloud of hot, highly ionised X-ray emitting gas.

\section{Conclusions\label{conclusions}}

So what is B1? On balance, we conclude that B1a and B1b are probably young
elliptical galaxies, perhaps analogues of E$+$A galaxies today. Both their
red colors and compact sizes are naturally explained by this model.
An extremely dusty starburst in the central regions of a disk galaxy
can however, with some fine tuning, reproduce the colors and sizes.
The bluer colors of the centres of both components, and the Ly$\alpha$
emission from the core of B1b are, however, arguments against the presence
of large amounts of dust in these systems.
Near-IR H-band spectroscopy on an 8m-class telescope should resolve
this ambiguity, by detecting the spectral signatures of a Balmer/4000\AA\
break. B1c appears to be a region of
relatively unobscured star formation, perhaps induced by the interaction
of B1a and B1b.

If B1a and B1b are elliptical galaxies, their
surface brightnesses are an order of magnitude greater than those that the
Kormendy relation would predict for galaxies of this size, and they show
strong radial color gradients, with blue cores. 
The very existence of massive elliptical galaxies at this redshift, if 
confirmed, is
interesting \citep[eg.][and references therein]{tre99}. These galaxies 
would give us a chance to study young elliptical galaxies for the first time.
Our observations would pose several puzzles:

\begin{itemize}

\item Why is the surface brightness of these galaxies so great? Is this
simply due to passive evolution?

\item Why do both galaxies show strong, but very similar color gradients?

\item Is the remarkable similarity and proximity of the two galaxies
coincidental, or is it telling us something about how cluster elliptical
galaxies form?

\item If these galaxies are so massive and red at $z = 2.38$, what would
they have looked like at $z \sim 3$ when they were forming?

\end{itemize}

The nature of the Ly$\alpha$ nebula is less well constrained by our
observations. The nebula could be excited by photoionization from
an AGN, by whatever powers low redshift cooling flow nebulae,
or by fast shocks powered by either superwinds or gravitational potential 
energy. Spatially resolved spectroscopy of other emission lines will help 
discriminate between these models.

\acknowledgments

We wish to thank Joss Bland-Hawthorn and Catherine Drake for their
assistance at with the TTF observations, and all the staff at CTIO
for heroic efforts to get the mosaic camera working for us. We are
grateful to Mike Dopita, Megan Donahue and Martijn de Kool and
Ian George for helpful discussions. Hsiao-Wen Chen was instrumental in 
getting the slitless STIS spectrum extraction software working. David
Hogg's cosmology crib sheet (astro-ph/9905116) was very helpful to us.





\end{document}